\documentclass{appolb}
\usepackage{graphicx}
\usepackage{amsmath,amstext,amsthm,amssymb,amsfonts,slashed}
\newcommand{\ii}{\ensuremath{\mathrm{i}}}
\newcommand{\dd}{\ensuremath{\mathrm{d}}}

\begin{document}
\title{Kinetics of the chiral phase transition \\ in a quark-meson $\sigma$
  model\thanks{Presented at XXIX$^{\text{th}}$ International Conference
    on Ultra-relativistic Nucleus-Nucleus Collisions (Quark Matter
    2022).}}

\author{Hendrik van Hees, Alex Meistrenko, Carsten Greiner
\address{Institut f{\"u}r Theoretische Physik, Johann Wolfgang Goethe-Universit{\"a}t,
Max-von-Laue-Str. 1, D-60438 Frankfurt am Main, Germany}
\address{Helmholtz Research Academy Hesse for FAIR, Campus Riedberg, Max-von-Laue-Str. 12, D-60438 Frankfurt am Main, Germany
}}
\maketitle
\begin{abstract}
  Using the two-particle irreducible (2PI) $\Phi$-functional formalism
  for self-consistent approximations of a linear-$\sigma$ model for
  quarks and mesons in and out of equilibrium, the build-up of
  fluctuations of net-baryon number during the time evolution of an
  expanding fireball is studied within a kinetic theory for the order
  parameter ($\sigma$ field) and quark distribution
  functions. Initializing the system with purely Gaussian fluctuations a
  fourth-order cumulant is temporarily built up due to the evolution of
  the $\sigma$-field. This is counterbalanced, however, by the
  dissipative evolution due to collisions between quarks, anti-quarks,
  mesons, and the mean field, depending on the speed of the fireball
  expansion.
\end{abstract}
  
\section{Introduction}

One important motivation for ultra-relativistic heavy-ion experiments,
as conducted, e.g., with the large-hadron collider at CERN, the
Relativistic Heavy-Ion Collider (RHIC) at BNL, and in the future at the
Facility for Antiproton and Ion Research (FAIR) is the understanding of
the phase diagram of strongly interacting matter under extreme
conditions of temperature and density. For small baryo-chemical
potentials, $\mu_{\text{B}}$, lattice-QCD calculations
\cite{FODOR200287,Aoki:2006we} show that the transition between a
quark-gluon plasma and a hadron-resonance gas as well as the chiral
transition is a smooth crossover at a transition temperature
$T_{\text{c}} \simeq 155 \, \text{MeV}$. Based on effective models like
the Nambu-Jona-Lasinio model, quark-meson models with constituent quarks
\cite{PhysRev.122.345,PhysRev.124.246,Jungnickel:1995fp,doi:10.1142/S0217751X03014034},
and their Polyakov-loop extended versions
\cite{PhysRevD.73.014019,PhysRevD.75.034007,PhysRevD.76.074023,HERBST201158},
at larger $\mu_{\text{B}}$ one expects a 1$^{\text{st}}$-order
transition line ending in a critical point with a 2$^{\text{nd}}$-order
transition
\cite{MASAYUKI1989668,PhysRevD.58.096007,doi:10.1142/S0217751X92001757,PhysRevD.75.085015}. The
main challenge is that this phase structure must be reconstructed from
the observables, which reflect the state of this medium at the end of
the fireball evolution (thermal freeze-out), which lasts only for very
short time at the order of some
$10\,\text{fm}/c \simeq 10^{-23} \, \text{s}$. A challenging theoretical
question therefore is, whether ``grand-canonical'' higher-order
cumulants of the net-baryon density can develop and survive the rapid
time evolution of the finite-size fireball, expected to occur when the
medium is undergoing a 1$^{\text{st}}$- or especially a
2$^{\text{nd}}$-order phase transition and whether corresponding
quantitative signatures of a possible critical point can be observed.

In this contribution we study this, employing a set of coupled equations
for the quarks, anti-quarks, and mesons as well as the order parameter,
$\sigma$, of the chiral symmetry within a linear quark-meson $\sigma$
model, derived from the two-particle irreducible functional ($\Phi$
functional) formalism \cite{Meistrenko:2020nwx}.

\section{The kinetic equations}

We start from an O(4) linear-$\sigma$ model for $\sigma$-mesons, pions,
and u- and d-quarks,
\begin{equation}
\begin{split}
  \mathcal L= &\sum_{i=1} \bar\psi_i \Big[\ii \partial\!\!\!/-g\left(\sigma+\ii
                \gamma_5\vec\pi\cdot\vec\tau\right)\Big]\psi_i \\
              &+\frac{1}{2}\left(\partial_\mu\sigma\partial^\mu\sigma
                +\partial_\mu\vec\pi
                \partial^\mu\vec\pi\right)-\frac{\lambda}{4}\left(\sigma^2+\vec\pi^2-\nu^2\right)^2+f_{\pi}m_\pi^2\sigma+U_0\, ,
\end{split}
\end{equation}
where $\lambda=20$, $f_{\pi}=93\,\text{MeV}$,
$m_{\pi}=138 \, \text{MeV}$, $\nu^2=f_{\pi}^2-m_{\pi}^2/\lambda$, and
$U_0=m_{\pi}^4/(4 \lambda) - f_{\pi^2} m_{\pi^2}$ are chosen to lead to
the right pion phenomenology in the vacuum. The quark-meson-coupling
constant $g$ is varied in the range between 2-5, leading to
cross-over as well as 1$^{\text{st}}$- and 2$^{\text{nd}}$-order chiral
phase transitions at finite $T$ and $\mu_{\text{B}}$.

\begin{figure}
\centerline{\includegraphics[width=0.47\linewidth]{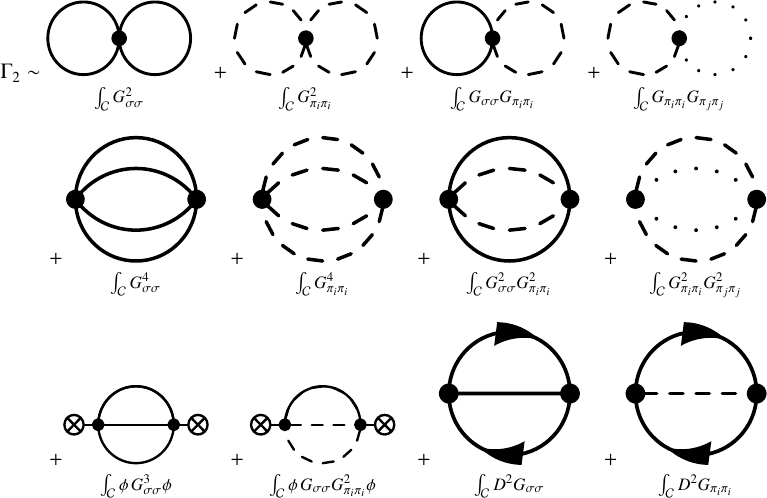}}
\caption{Included 2PI part of the effective action with 1$^{\text{st}}$ line:
  Hartree diagrams, 2$^{\text{nd}}$ line: basketball diagrams, 3$^{\text{rd}}$ line: sunset
  diagrams, where solid lines stand for the $\sigma$ propagator, dashed
  and pointed lines for the pion propagators and solid lines with arrows
  for the fermion propagator. The circle with a cross represents a
  $\sigma$ mean-field.}
\label{fig.1}
\end{figure}

The kinetic equations to describe both the equilibrium state as well as
the off-equilibrium kinetic evolution of this model we use the 2PI
$\Phi$-derivable approximation, defined in terms of the corresponding Feynman
diagrams in Fig.\ \ref{fig.1}.
Solving the corresponding self-consistent equations for the propagators
and the mean $\sigma$-field in thermal equilibrium indeed leads to a
phase diagram with a cross-over transition at lower $\mu_{\text{B}}$ and
a first-order transition line ending in a critical point at
$(T,\mu_{\text{B}})=(108,157)\,\text{MeV}$ (for a quark-meson coupling,
$g=3.3$).

\begin{figure}
\centerline{\includegraphics[width=0.7\linewidth]{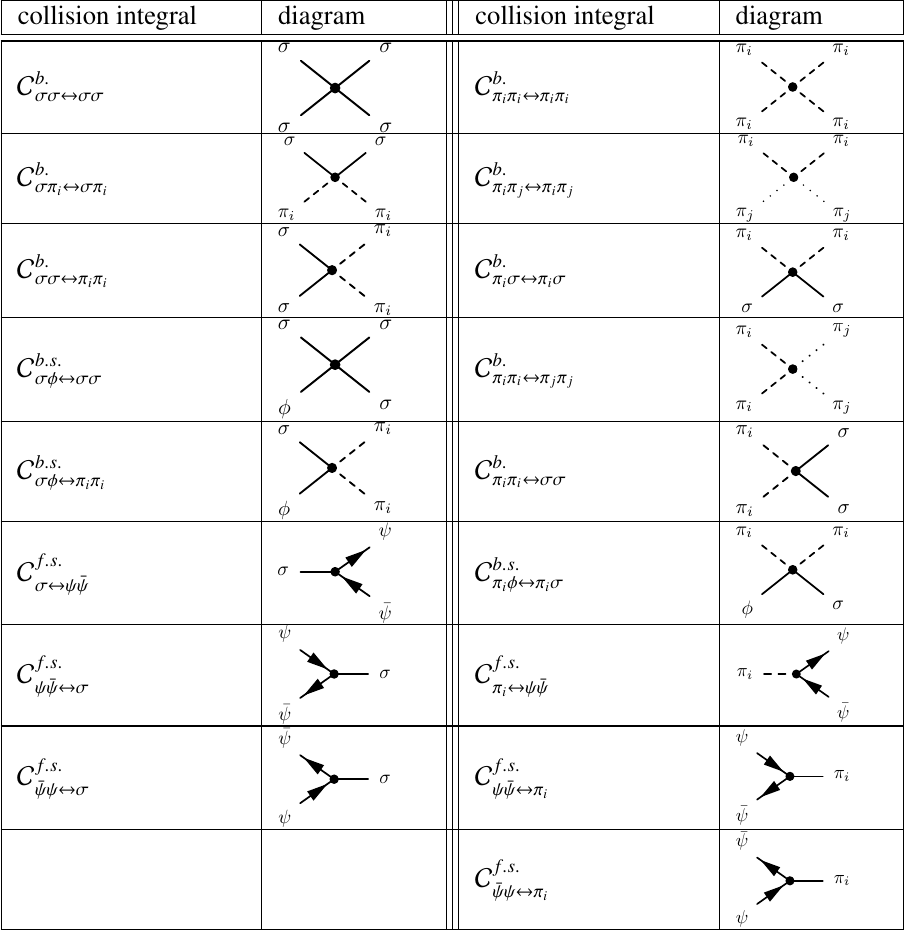}}
\caption{The scattering processes in the collision integrals of the
  kinetic equation. A full $\sigma$-line indicates a mean-field
  contribution, while a full $\phi$-line describes a scattering process
  involving a $\sigma$ meson.}
\label{fig.2}
\end{figure}

For the derivation of coupled kinetic equations of motion for the mean
$\sigma$-field and the generalized Boltzmann equations for the quark-
and meson-phase-space-distribution functions the diagrams are evaluated
within the Schwinger-Keldysh real-time formalism, leading to
corresponding Kadanoff-Baym equations. Then a first-order
gradient-expansion approximation to the Wigner transforms of the Green's
functions as well as an ``on-shell approximation'' with self-consistent
dispersion relations has been applied. This results in a non-Markovian
dissipative equation for the mean field, $\sigma$, and a Boltzmann
equation with a collision integral including the scattering processes
depicted in Fig.\ \ref{fig.2}.

\section{Simulation of a heavy-ion collision}

To simulate the formation of higher-order cumulants of
net-baryon-density fluctuations in momentum bins, we describe the
fireball of strongly interacting quark-meson matter by an expanding
homogeneous and isotropic Friedmann-Lema{\^i}tre-Robertson-Walker
metric, $\dd s^2 = \dd t^2 -a^2(t) (\dd x_1^2+\dd x_2^2+\dd x_3^2)$.
This only leads to a modification in the drift terms of the mean-field
and kinetic equations. For the mean field the ``Hubble expansion'' adds
an additional dissipation term $3 H \partial_t \sigma$ with the ``Hubble
constant'' $H=\dot{a}/a$, and functions an additional term of the form
$-H p \partial_p f(t,p)$ in the drift terms for the particle phase-space
distribution. Note that due to the assumed spatial homogeneity and
isotropy the $f$'s only depend on $t$ and $p=|\vec{p}|$.

To initialize the fireball a spherically symmetric bubble of radius
$R_0=5 \; \text{fm}$ is considered, which then is expanding according to
the above defined FLRW expansion with $a=v t$. The medium within this
bubble is initialized in thermal equilibrium with a temperature $T_0$
and baryon chemical potential $\mu_{\text{B}0}$. Then the initial
net-quark number is Monte-Carlo sampled corresponding to a Gaussian
distribution with the mean determined by the thermal initial state and a
standard deviation of
$\sigma_{q,\text{net}}=\langle N_{q,\text{net}} \rangle/10$. To mimick
the expected fluctuations in a heavy-ion collision within a given
``centrality bin'' we keep the parameters $R_0$ and $T_0$ fixed and
adjust $\mu_{q}$ such that the fireball contains the net-quark number
$N_{q,\text{net}}$ specified by the Monte-Carlo sampling.

With this initial conditions the coupled mean-field and kinetic
integro-differential equations of motion are solved numerically on a
momentum grid. It has been checked that the total net-quark
number is conserved within a few precent numerical accuracy. 

\begin{figure}
\centerline{\includegraphics[width=0.98\linewidth]{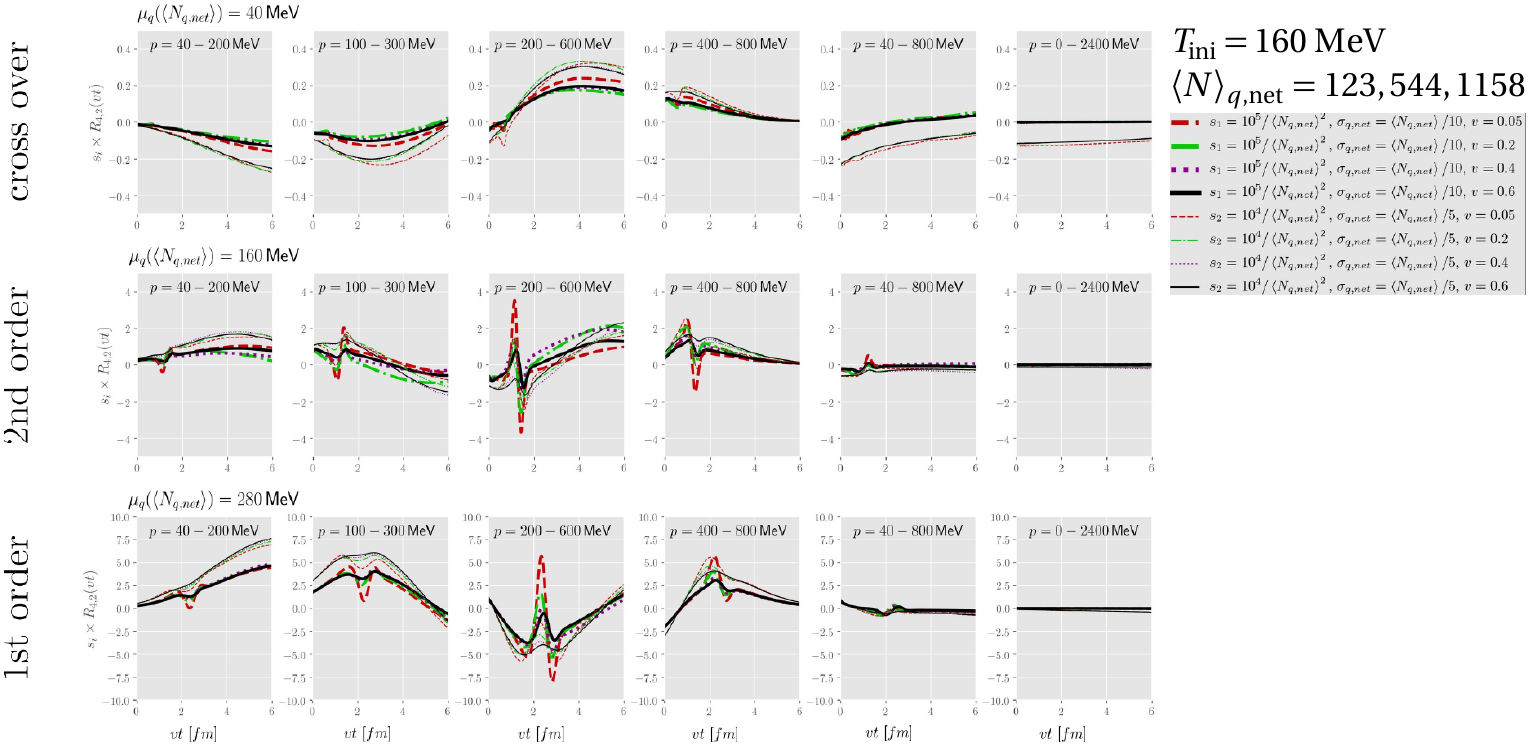}}
\caption{Results for the rescaled cumulant ratio $R_{4,2}$ for different
  initial conditions where the fireball evolves through a cross-over,
  second-order, or first-order transition.}
\label{fig.3}
\end{figure}

In Fig.\ \ref{fig.3} we show the results for the cumulant ratio,
$R_{4,2}=\kappa_4/\kappa_2$, for initial conditions adjusted such that
the system undergoes cross-over, second-order, and first-order
transition, respectively.
The fluctuations are plotted in different momentum intervals and for
different expansion velocities, $v$, as a function of $v t$. I turns out
that the most pronounced fluctuations occur at the critical time scales
$\tau_{m_{\sigma},\text{min}}$ (dynamical minimum of the $\sigma$ mass)
and $\tau_{\sigma \rightarrow q \bar{q}}$ ($q\bar{q}$-pair production
from $\sigma$ decay). The fluctuations become largest for the smallest
expansion velocity of $v=0.05 c$, corresponding to a quasi-adiabatic
expansion, where the system stays for the longest time close to the
critical region. However, in relativistic heavy-ion collisions this
intermediate build-up of fluctuations related with the critical region
of the phase diagram cannot be observed but only those surviving until
the thermal freeze-out, which corresponds in our model to
$v t \geq 6 \;\ \text{fm}$ and a fireball radius of
$R \geq 11 \; \text{fm}$.

In the most interesting case, $\mu_q=160 \; \text{MeV}$, where the
system evolves close to the critical point of a 2$^{\text{nd}}$-order
phase transition the largest cumulant ratio in the final state is
observed for intermediate expansion velocities $v=0.2$-$0.4 c$, while
for the case when the system goes through a 1$^{\text{st}}$-order phase
transition the final fluctuations are rather insensitive to the
expansion velocity (for the most interesting momentum range
$p=200$-$600 \; \text{meV}$). This allows in principle to distinguish
between different types of the phase transition and indicates the
expected longer relaxation times (``critical slowing down'') around a
critical point in the phase diagram, i.e., the system needs longer to
equilibrate and thus the fluctuations survive until the thermal
freeze-out. The absolute magnitude of the cumulant ratio increases with
an increasing net-baryon number (note the scaling factors
$s_1,s_2 \sim 1/\langle N_{q,\text{net}} \rangle^2$ in the plots of
Fig.\ \ref{fig.3}).

\section{Conclusions}

Although the fluctuations of net-baryon numbers in an expanding finite
system are less pronounced compared to the expectations from a
equilibrated infinite strongly interaction matter, our simulations
suggest that a significant deviation from the crossover behavior is
observable through higher-order cumulant ratios in different momentum
bins, providing a positive candidate for an experimental signature of
the chiral phase transition and a possible critical region in the phase
diagram of strongly interacting matter.


\begin{thebibliography}{10}
\providecommand{\url}[1]{\texttt{#1}}
\providecommand{\urlprefix}{}
\providecommand{\eprint}[2][]{\url{#2}}

\bibitem{FODOR200287}
Z.~Fodor and S.~Katz, Physics Letters B \textbf{534}, 87  (2002),
  \urlprefix\url{https://doi.org/10.1016/S0370-2693(02)01583-6}.

\bibitem{Aoki:2006we}
Y.~Aoki, G.~Endrodi, Z.~Fodor, S.~D. Katz, and K.~K. Szabo, Nature
  \textbf{443}, 675 (2006),
  \urlprefix\url{https://dx.doi.org/10.1038/nature05120}.

\bibitem{PhysRev.122.345}
Y.~Nambu and G.~Jona-Lasinio, Phys. Rev. \textbf{122}, 345 (1961),
  \urlprefix\url{https://dx.doi.org/10.1103/PhysRev.122.345}.

\bibitem{PhysRev.124.246}
Y.~Nambu and G.~Jona-Lasinio, Phys. Rev. \textbf{124}, 246 (1961),
  \urlprefix\url{https://dx.doi.org/10.1103/PhysRev.124.246}.

\bibitem{Jungnickel:1995fp}
D.~U. Jungnickel and C.~Wetterich, Phys. Rev. D \textbf{53}, 5142 (1996).

\bibitem{doi:10.1142/S0217751X03014034}
J.~Berges, D.-U. Jungnickel, and C.~Wetterich, International Journal of Modern
  Physics A \textbf{18}, 3189 (2003),
  \urlprefix\url{https://dx.doi.org/10.1142/S0217751X03014034}.

\bibitem{PhysRevD.73.014019}
C.~Ratti, M.~A. Thaler, and W.~Weise, Phys. Rev. D \textbf{73}, 014019 (2006),
  \urlprefix\url{https://dx.doi.org/10.1103/PhysRevD.73.014019}.

\bibitem{PhysRevD.75.034007}
S.~R\"o\ss{}ner, C.~Ratti, and W.~Weise, Phys. Rev. D \textbf{75}, 034007
  (2007), \urlprefix\url{https://dx.doi.org/10.1103/PhysRevD.75.034007}.

\bibitem{PhysRevD.76.074023}
B.-J. Schaefer, J.~M. Pawlowski, and J.~Wambach, Phys. Rev. D \textbf{76},
  074023 (2007), \urlprefix\url{https://dx.doi.org/10.1103/PhysRevD.76.074023}.

\bibitem{HERBST201158}
T.~K. Herbst, J.~M. Pawlowski, and B.-J. Schaefer, Physics Letters B
  \textbf{696}, 58  (2011),
  \urlprefix\url{https://doi.org/10.1016/j.physletb.2010.12.003}.

\bibitem{MASAYUKI1989668}
A.~Masayuki and Y.~Koichi, Nuclear Physics A \textbf{504}, 668  (1989),
  \urlprefix\url{https://doi.org/10.1016/0375-9474(89)90002-X}.

\bibitem{PhysRevD.58.096007}
M.~A. Halasz, A.~D. Jackson, R.~E. Shrock, M.~A. Stephanov, and J.~J.~M.
  Verbaarschot, Phys. Rev. D \textbf{58}, 096007 (1998),
  \urlprefix\url{https://dx.doi.org/10.1103/PhysRevD.58.096007}.

\bibitem{doi:10.1142/S0217751X92001757}
F.~Wilczek, International Journal of Modern Physics A \textbf{07}, 3911 (1992),
  \urlprefix\url{https://dx.doi.org/10.1142/S0217751X92001757}.

\bibitem{PhysRevD.75.085015}
B.-J. Schaefer and J.~Wambach, Phys. Rev. D \textbf{75}, 085015 (2007),
  \urlprefix\url{https://dx.doi.org/10.1103/PhysRevD.75.085015}.

\bibitem{Meistrenko:2020nwx}
A.~Meistrenko, H.~van Hees, and C.~Greiner, Annals Phys. \textbf{431}, 168555
  (2021), \urlprefix\url{https://doi.org/10.1016/j.aop.2021.168555}.

\end{thebibliography}

\end{document}